\documentstyle[12pt]{article}
\let\ssection=\section
\renewcommand{\section}{\setcounter{equation}{0}\ssection}

\newcommand\mathC{\mkern1mu\raise2.2pt\hbox{$\scriptscriptstyle|$}
        {\mkern-7mu\rm C}}          
\newcommand{\mathR}{{\rm I\! R}}         
\newcommand{\mathZ}{{\rm Z\!\! Z}}
\newcommand\Q{{\cal Q}}

\newcommand\K{{\cal K}}
\newcommand\id{{\rm id}}
\newcommand\mathN{{\rm I\! N}}

\newcommand\braket[2]{\langle#1|#2\rangle}

\newcommand\Ob[1]{{\rm Ob}(#1)}
\newcommand\Hom[1]{{\rm Hom}(#1)}
\newcommand\Map[1]{{\rm Map}(#1)}

\newcommand\Ran[1]{{\rm Ran}\;#1}
\newcommand\Dom[1]{{\rm Dom}\;#1}
\newcommand\AF[1]{{\rm AF}(#1)}
\newcommand\eq[1]{Eq.\ (\ref{#1})}

\newcommand\longvec[1]{\overline{#1}}

\newcommand\mapdown[1]{\Big\downarrow
        \rlap{$\vcenter{\hbox{$\scriptstyle#1$}}$}}

\newcommand\shortmapright[1]{\smash{
        \mathop{\mbox{\large{$\;\rightarrow\;$}}}\limits^{#1}}}
\newcommand\mapleft[1]{\smash{
        \mathop{\mbox{\large{$\;\longleftarrow\;$}}}\limits^{#1}}}
\newcommand\bundle[3]{\begin{array}[t]{c}
        {#1}\\ \mapdown{#2}\\ {#3}\end{array}}

\newcommand\bundlemapright[2]{\begin{array}[t]{c}
\mapleft{#1}\\
\phantom{\mapdown{}}\\\mapleft{#2}\\\end{array}}

\begin{document}
\begin{titlepage}
\hspace{09truecm}Imperial/TP/2-03/11

\hspace{09truecm}gr-qc/0304077

\begin{center}
{\large\bf A New Approach to Quantising Systems:\\[6pt]
        II.\ Quantising on a Category of Sets}
\end{center}

\vspace{0.8 truecm}

\begin{center}
            C.J.~Isham\footnote{email: c.isham@imperial.ac.uk}\\[10pt]
            The Blackett Laboratory\\
            Imperial College of Science, Technology \& Medicine\\
            South Kensington\\
            London SW7 2BZ\\
\end{center}

\begin{abstract}

In \cite{IshQCT1_03}, a new approach was suggested for finding
quantum structures that might, in particular, be used in potential
approaches to quantum gravity that involve non-manifold models for
space and/or space-time (for example, causal sets). This involved
developing a procedure for quantising a system whose configuration
space---or history-theory analogue---is the set of objects in a
(small) category $\Q$. In the present paper, we show how this
theory can be applied to the special case when $\Q$ is a category
of sets.

\end{abstract}
\end{titlepage}

\section{Introduction}\label{Sec:Introduction}
In \cite{IshQCT1_03}, hereafter referred to as {\bf I.}, a new
approach was developed for constructing quantum theories that
might, in particular, be used in potential approaches to quantum
gravity that involve non-manifold models for space and/or
spacetime. This could include things like `quantum topology',
`quantum manifolds', `quantum causal sets' {\em etc\/}.

The starting point is the observation that, in each such example,
the entities of interest are the objects of a {\em category\/}
whose arrows are structure-preserving maps. This suggested
tackling the more general problem of finding a quantum theory of a
system whose configuration space (or history-theory analogue) is
the set of objects $\Ob\Q$ of an arbitrary small category $\Q$. A
general solution to this problem of `quantising on a category' was
presented in {\bf I.}, and in the present paper we show how this
theory works in practice for categories of sets.

First, a summary of the results obtained in {\bf I}. The
methods used were motivated by considering, as an
analogy, a system whose configuration spaces is a
homogeneous space $Q\simeq G/H$, where $G$ and $H$ are
Lie groups. In particular, the analogue of the group $G$
is chosen to be the set $\AF\Q$ of `arrow fields' on
$\Q$, where an {\em arrow field\/} is defined to be a map
$X:\Ob\Q\rightarrow\Hom\Q$ (the arrows of $\Q$) such
that, for each $A\in\Ob\Q$, the domain of $X(A)$ is $A$;
thus $X(A):A\rightarrow B$ for some $B\in\Ob\Q$. A
simple, and useful, example of an arrow field is when all
but one of the arrows $X(A)$, $A\in\Ob\Q$, is the
identity $\id_A:A\rightarrow A$. Specifically, for each
arrow $f$, an arrow field $X_f$ can be defined by
\begin{equation}
    X_f(A):=\left\{ \begin{array}{ll}
            f &\mbox{\ if $\Dom f=A$;}
                                    \label{Def:Xf}\\
                 \id_A & \mbox{\ otherwise,}
                 \end{array}
        \right.
\end{equation}
for all $A\in\Ob\Q$.

A crucial property of arrow fields is that they form a
monoid\footnote{A monoid is a semigroup with a unit
element. In the present case, the unit element is the
arrow field $\iota$ defined by $\iota(A):=\id_A$ for all
$A\in\Ob\Q$.}. Specifically: if $X_1,X_2\in\AF\Q$, we
construct an arrow field $X_2\&X_1$ by defining the arrow
$(X_2\&X_1)(A)$, $A\in\Ob\Q$, to be the composition of
the arrow $X_1(A)$ with the arrow obtained by evaluating
$X_2$ on the range of $X_1(A)$:
\begin{equation}
    (X_2\&X_1)(A):=X_2(\Ran{X_1(A)})\circ X_1(A).
\end{equation}
Put more simply, if $X_1(A):A\rightarrow B$, then
$(X_2\&X_1)(A):=X_2(B)\circ X_1(A)$.

An action of the monoid $\AF\Q$ on $\Ob\Q$ can be defined
by $\ell_X(A):=\Ran X(A) $, so that if $X(A):A\rightarrow
B$ then $\ell_X(A)=B$. This action was used in {\bf I.}
to construct an operator (anti-) representation of
$\AF\Q$ on complex-valued functions on $\Ob\Q$ by
defining, for all $A\in\Ob\Q$,
\begin{equation}
        (\hat a(X)\psi)(A):=\psi(\ell_XA).
        \label{Def:aX}
\end{equation}
These operators satisfy
\begin{equation}
        \hat a(X_2)\hat a(X_1)=\hat a(X_1\&X_2)
\end{equation}
for all arrow fields $X_1$,$X_2$. If \eq{Def:aX} is applied to the
special arrow fields $X_f$ in \eq{Def:Xf} then, defining $\hat
a(f):=\hat a(X_f)$, we get
\begin{equation}
    (\hat a(f)\psi)(A)=\left\{ \begin{array}{ll}
                \psi(\Ran f)&\mbox{\ if $\Dom f=A$;}
                \label{aXf}\\[3pt]
                 \psi(A) & \mbox{\ otherwise.}
                 \end{array}
        \right.
\end{equation}

The inner product on the quantum state functions is
\begin{equation}
    \braket\phi\psi:=\int_{\Ob\Q}d\mu(A)\,\phi^*(A)\psi(A)
                \label{ScalProdMeas}
\end{equation}
for some measure $\mu$ on $\Ob\Q$. If $\Ob\Q$ is finite, or
countably infinite, a natural choice for the inner product is
\begin{equation}
        \braket\phi\psi:=\sum_{A\in\Ob\Q}\phi(A)^*\psi(A).
            \label{ScalProdCount}
\end{equation}

The `configuration' space variables are members of the
space $F(\Ob\Q,\mathR)$ of all real-valued functions on
$\Ob\Q$, and unitary operators $\hat V(\beta)$, $\beta\in
F(\Ob\Q,\mathR)$, are defined by
\begin{equation}
        (\hat V(\beta)\psi)(A):=e^{-i\beta(A)}\psi(A)
        \label{Def:V(beta)}
\end{equation}
for all $A\in\Ob\Q$.

The operators $\hat a(X)$, $X\in\AF\Q$, and $\hat
V(\beta)$, $\beta\in F(\Ob\Q,\mathR)$, satisfy the
relations
\begin{eqnarray}
    \hat a(X_2)\hat a(X_1)&=&\hat a(X_1\&X_2)\label{CQAaa}\\
    \hat V(\beta_1)\hat V(\beta_2)&=&\hat
    V(\beta_1+\beta_2)                  \label{CQAVV}\\
    \hat a(X)\hat V(\beta)&=&\hat V(\beta\circ\ell_X)\hat
    a(X)                                \label{CQAaV}
\end{eqnarray}
and, as such, constitute a representation of the
`category quantisation monoid', which is defined to be
the semi-direct product $\AF\Q\times_\ell
F(\Ob\Q,\mathR)$. In general, the possible quantisations
of the system are deemed to be given by the irreducible,
faithful representations of this monoid.

The representation in \eq{Def:aX} and \eq{Def:V(beta)} is
inadequate because it fails to separate
arrows\footnote{Two arrows $f,g$ with the same domain and
range are said to be{\em separated\/} in the quantum
theory if $\hat a(X_f)\neq \hat a(X_g)$.} with the same
domain and range. This can be rectified by introducing a
suitable bundle of Hilbert spaces, $A\leadsto \K[A]$,
with linear `multipliers'
$m(X,A):\K[\ell_XA]\rightarrow\K[A]$. The arrow-field
operators $\hat a(X)$, $X\in\AF\Q$, are now defined as
\begin{equation}
    (\hat a(X)\psi)(A):=m(X,A)\psi(\ell_XA)
    \label{Def:a(X)psi}
\end{equation}
and satisfy \eq{CQAaa} provided the multipliers are subject
to the conditions
\begin{equation}
    m(Y\&X,A)=m(X,A)m(Y,\ell_XA)   \label{multconds}
\end{equation}
for all $X,Y\in\AF\Q$ and $A\in\Ob\Q$. Note that if the
arrow fields $X,Y$ are such that, for some $A$,
$X(A):A\rightarrow A$ and $Y(A):A\rightarrow A$, then
\eq{multconds} gives
\begin{equation}
            m(Y\&X,A)=m(X,A)m(Y,A)  \label{RepHom(AA)}
\end{equation}
which corresponds to an anti-representation of the monoid
$\Hom{A,A}$ on the Hilbert space $\K[A]$.

As explained in {\bf I.}, a natural way of generating
structures of this type is to start with a presheaf of
Hilbert spaces over $\Ob\Q$, so that any arrow
$f:A\rightarrow B$ is `covered' by a linear map
$\kappa(f):\K[B]\rightarrow \K[A]$, and then to define a
multiplier by $m(X,A):=\kappa(X(A))$ for all $X\in\AF\Q$,
$A\in\Ob\Q$. The quantum states are defined to be
cross-sections of the Hilbert bundle, with the inner
product
\begin{equation}
    \braket\phi\psi:=
        \int_{\Ob\Q}d\mu(A)\,\langle\phi(A),\psi(A)\rangle_{\K[A]}
                \label{Def:innerproduct}
\end{equation}
where $\langle\,\cdot,\cdot\,\rangle_{\K[A]}$ denotes the
inner product in the Hilbert space $\K[A]$, and $\mu$ is
a measure on $\Ob\Q$. The operators $\hat V(\beta)$ are
defined as in \eq{Def:V(beta)}.

The aim of the present paper is to give some examples of
this scheme to illustrate the basic ideas. We start in
Section \ref{Sec:SimplestExamples} with the simple case
when $\Q$ is a partially-ordered set whose objects are
points with no internal structure, and multipliers are
not needed. The heart of the paper is Section
\ref{Sec:QST} which contains a discussion of the
quantisation on a category of sets. Section
\ref{SubSec:QSTFinite} focusses on a category of finite
sets, and we show how to construct explicit multiplier
representations that separate the arrows in the category.
In Section \ref{SubSec:QMT} we discuss briefly how these
methods can be extended to non-finite sets with the aid
of categories of measure spaces. Then, after a few
cautionary remarks about adapting the scheme to
categories of causal sets, we give a complete
quantisation in Section \ref{SubSec:VSEx} of a very
simple category with just two causal sets. In Section
\ref{Sec:FurtherQuOps} we show how certain special
operations on causal sets---for example, taking the
linear sum of two of them---can be incorporated naturally
into the scheme, as can the lattice operations on a
lattice. Finally, in Section \ref{Sec:QFT} we return to a
poset category to show how standard quantum field theory
is another special example of the scheme.

\section{Some Simple Examples}
\label{Sec:SimplestExamples}
\subsection{The Simplest Example}
The simplest example is when $\Q$ is a monoid $M$
considered as a category. There is just one object $\#$,
and the arrows are in one-to-one correspondence with the
elements of $M$, with arrow composition being monoid
multiplication. An arrow field is simply an arrow; {\em
i.e.}, an element of $M$. A quantum theory on this
category can be found by associating with $\#$ the
Hilbert space $\K$ of any faithful, irreducible
representation $m\mapsto \hat R(m)$ of $M$. The state
vectors are functions from $\#$ to $\K$, and are hence
just elements of $\K$. If $m$ is an
arrow-field/arrow/monoid element, then $\hat a(m):=\hat R
(m)$. The faithfulness of the representation guarantees
that it separates the arrows, {\em i.e.}, the elements of
$M$.

\subsection{A Poset Category}
\label{SubSec:PosetCategory} A monoid considered as a
category has only one object, but many arrows. At the
other extreme is a partially ordered set (poset) $P$
considered as a category. The objects are the elements of
$P$, and there is an arrow $o_{p,q}:p\rightarrow q$ if,
and only if, $p\leq q$. Thus there is at most one arrow
between each pair of objects---in particular,
$\Hom{p,p}=\{\id_p\}$ for all $p\in P$---but many
objects.

It is important to be clear about the physical situations
that could be represented by such a model. For example,
consider a causal set theory of the type discussed
recently by Dowker\footnote{Fay Dowker, private
communication.} in which, in the language of the present
paper, the objects of the category $\Q$ are subsets of a
single, `master' causal set $\Gamma$. Superficially, this
category might look like a poset, with the ordering being
subset inclusion. However, this is misleading since each
object $A\subset\Gamma$ has an internal structure, as
reflected in the monoid $\Hom{A,A}$ of order-preserving
maps from $A$ to itself, and one would want to represent
this in the quantum theory. More generally, if $A$ and
$B$ are any non-trivial causal subsets of $\Gamma$ with
$A\subset B$, then $\Hom{A,B}$ typically contains more
than just the subset inclusion map $i_{A,B}:A\subset B$.
For example, if $\alpha\in\Hom{A,A}$ and
$\beta\in\Hom{B,B}$, then
\begin{equation}
    \beta\circ i_{A,B}\circ\alpha:A\rightarrow B\label{bia}
\end{equation}
belongs to $\Hom{A,B}$. Thus $\Q$ is not just a poset
category.

When we do have a genuine poset category $P$, there is at
most one arrow between any pair of objects, which means
that to separate arrows in the quantum theory it suffices
to let the state space be just complex-valued functions
on $P$. Then, if $p\leq q$, the operator $\hat
a(p,q):=\hat a(X_{o_{p,q}})$ associated with the arrow
$o_{p,q}:p\rightarrow q$ is (see \eq{aXf})
\begin{equation}
        (\hat a(p,q)\psi)(r)=\left\{ \begin{array}{ll}
            \psi(q) &\mbox{\ if $p=r$;}
            \label{Def:a(pq)}\\[3pt]
                 \psi(r) & \mbox{\ otherwise.}
                 \end{array}
        \right.
\end{equation}
As in \eq{Def:V(beta)}, functions
$\beta:P\rightarrow\mathR$ are represented  by
\begin{equation}
        (\hat V(\beta)\psi)(r):=e^{-i\beta(r)}\psi(r)
\end{equation}
for all $r\in P$. The natural inner product is the one  in
\eq{ScalProdCount} provided $P$ has at most a countable
number of elements. For a system whose configuration space
(or history-theory analogue) is a poset with a finite
number, $n$, of elements, the quantum state space is just
$\mathC^n$.

\subsection{Quantum Theory on $\mathN$}
A simple example of a poset is a finite chain with the
Hasse diagram
\begin{equation}
0\rightarrow 1\rightarrow 2 \rightarrow\cdots\rightarrow
n-1     \label{HasseDiagramn}
\end{equation}
in which the only arrows between objects/points are those
shown in \eq{HasseDiagramn}, plus combinations of them,
plus the identity arrow at each object. The quantisation
goes ahead as above, and the Hilbert state space is
$\mathC^n$.

More interesting is the generalisation with a countably
infinite number of objects
\begin{equation}
    0\rightarrow 1 \rightarrow 2 \rightarrow\cdots
    \rightarrow n-1 \rightarrow\cdots\label{DiagramN}
\end{equation}
so that $\Ob\Q$ is the natural numbers $\mathN$. In this
case, there is a special family of arrow fields $X^m$,
$m\in\mathN$, defined on all $n\in\mathN$ by
\begin{equation}
            X^{m}(n):=n\rightarrow (n+m)
\end{equation}
where the right hand side denotes the combination of $m$
successive arrows in the Hasse diagram of \eq{DiagramN}.
Clearly
\begin{equation}
        X^{m_1}\&X^{m_2}=X^{m_1+m_2},
\end{equation}
and hence the set of all arrow fields of this type is a
submonoid of $\AF\Q$. It is isomorphic to the additive
monoid $\mathN$.

This submonoid acts on state functions as
\begin{equation}
            (\hat a(X^m)\psi)(n)=\psi(n+m),\label{Def:aXm}
\end{equation}
for all $n\in\mathN$. Together with the representation
\eq{Def:V(beta)} of the space of functions
$F(\mathN,\mathR)$, this gives an irreducible
representation of the semi-direct product monoid
$\mathN\times_\ell F(\mathN,\mathR)$. This is a submonoid
of the full category quantisation monoid
$\AF\Q\times_\ell F(\mathN,\mathR)$. This is an example
of the general fact that for an arbitrary small category
$\Q$ there may well be submonoids of $\AF\Q\times_\ell
F(\Ob\Q,\mathR)$ for which an irreducible representation
of $\AF\Q\times_\ell F(\Ob\Q,\mathR)$ remains
irreducible. This is analogous to what happens when
$Q\simeq G/H$, where an irreducible representation of the
semi-direct product ${\rm Diff}(Q)\times_d
C^\infty(Q,\mathR)$ remains irreducible when restricted
to the subgroup $G\times_\tau W$, where $W$ is a special
finite-dimensional subspace of $C^\infty(Q,\mathR)$ (for
more details see {\bf I}).

Note that the  example in \eq{Def:aXm} illustrates
something that is missing for a general small category:
namely, the existence of arrow fields that give the
`same' arrow at each object.

We could write down what is superficially the same
category as in \eq{DiagramN} but with the object $n$ now
being identified with the set $\{0,1,2,\ldots,n-1\}$.
Thus the Hasse diagram of \eq{DiagramN} is replaced with
\begin{equation}
    \{0\}\rightarrow
    \{0,1\}\rightarrow\{0,1,2\}\rightarrow \ldots
    \label{Ncategoryplus}
\end{equation}
where the arrow between each pair of objects is now
interpreted as a subset embedding. This gives rise to a
quantum theory of finite sets in which the number of
elements in a set can change.

However, the situation here is essentially the same as
that mentioned in Section \ref{SubSec:PosetCategory} in
the context of Fay Dowker's model. Specifically, if each
object $A$ is viewed as an explicit set
$\{0,1,\ldots,...,n-1\}$, then it now has an internal
structure, as reflected in the non-triviality of
$\Hom{A,A}$. Thus the category is different from the
starting example in \eq{DiagramN}, and the quantisation
should reflect this. In particular, the monoids
$\Hom{A,A}$ should be represented faithfully, and hence
separate arrows of the type $\beta\circ
i_{A,B}\circ\alpha:A\rightarrow B$, where
$\beta:B\rightarrow B$, $\alpha: A\rightarrow A$, and
$i_{A,B}:A\rightarrow B$ is the concatenation of the
appropriate number of embeddings in \eq{Ncategoryplus}.
This requires the introduction of non-trivial Hilbert
spaces $\K[A]$ and associated multipliers.

\section{Quantising on a Category of Sets}
\label{Sec:QST}
\subsection{The Example Of Finite Sets}\label{SubSec:QSTFinite}
\subsubsection{Constructing the Multipliers}
\label{SubSubSec:QSTMultipliersFinite} The physical examples in
which we are most interested arise when $\Q$ is a small category
of causal sets ({\em i.e.}, posets), or topological spaces, or
differentiable manifolds, or any other such mathematical entities
that are potential models for space-time, or space. The objects in
these categories are sets with structure, and the arrows are
structure-preserving maps. Here we shall discuss in detail the
most basic situation of this type, which is when $\Q$ is a small
category of sets, and the arrows are {\em any\/} maps between a
pair of sets. The resulting theory can then be used as a stepping
stone for handling categories whose objects have a more specific
structure, and whose morphisms are therefore more restricted.

The operators $\hat a(X)$ that represent arrow fields are
defined in \eq{Def:a(X)psi}, and---as in {\bf I.}---we
shall assume that the multiplier
$m(X,A):\K[\ell_XA]\rightarrow\K[A]$ is obtained from a
family of linear maps $\kappa(f):\K[\Ran f]
\rightarrow\K[\Dom f]$, $f\in\Hom\Q$, such that
$m(X,A)=\kappa(X(A))$ for all $X\in\AF\Q$ and
$A\in\Ob\Q$.  In order to ensure the multiplier
conditions \eq{RepHom(AA)}, these maps must satisfy the
coherence conditions
\begin{equation}
\kappa(g\circ f)=\kappa(f)\kappa(g):\K[C]\rightarrow
\K[A]\label{lambdaConditions}
\end{equation}
for any $f:A\rightarrow B$ and $g:B\rightarrow C$. As
discussed in {\bf I.}, the conditions
\eq{lambdaConditions} are equivalent to the existence of
a presheaf of Hilbert spaces on $\Q$, where a function
$f:A\rightarrow B$ is `covered' by the linear map
$\kappa(f):\K[B]\rightarrow \K[A]$.

A necessary condition for the quantum theory to separate
arrows is that each Hilbert space $\K[A]$, $A\in\Ob\Q$,
must carry a faithful representation of the monoid
$\Map{A,A}$ ($=\Hom{A,A}$) of all maps from the set $A$
to itself. It would be very desirable if we could find a
suitable presheaf that is {\em generic\/} in the sense
that it can be defined for any small category of sets
without invoking the specific details of the category.

One vector space that is naturally associated with any
set $A$, is the space $F(A,\mathC)$ of complex-valued
functions on $A$. Then, to each arrow/function
$f:A\rightarrow B$, there corresponds an obvious linear
map $\kappa(f):F(B,\mathC)\rightarrow F(A,\mathC)$
defined on $v\in F(B,\mathC)$ by
\begin{equation}
    (\kappa(f)v)(a):=v(f(a))       \label{Def:lfva}
\end{equation}
for all $a\in A$.

To check the coherence conditions \eq{lambdaConditions},
suppose $f:A\rightarrow B$ and $g:B\rightarrow C$, and let
$v\in F(C,\mathC)$. Then
\begin{equation}
    (\kappa(f)\kappa(g)v)[a]=(\kappa(g)v)[f(a)]=v[g(f(a)]
                \label{lflgv}
\end{equation}
However, $g(f(a))=g\circ f(a)$, and so \eq{lflgv} gives
\begin{equation}
(\kappa(f)\kappa(g)v)[a]=v[g\circ f(a)]=(\kappa(g\circ
f)v)[a]
\end{equation}
and hence $\kappa(f)\kappa(g)=\kappa(g\circ f)$, as
required.

As it stands, $F(A,\mathC)$ is merely a complex vector
space. However, if $A$ is a {\em finite\/} set, there is
a vector space isomorphism $F(A,\mathC)\simeq
\mathC^{|A|}$, where $|A|$ denotes the number of elements
in $A$. The standard inner product on $\mathC^{|A|}$ can
then be used to equip $F(A,\mathC)$ with an inner
product.  In \cite{IshQCT3_03}, this way of quantising on
a category of finite sets is rederived using a different
technique based on taking state vectors as complex-valued
functions on the set of arrows $\Hom{\Q}$.

If $A$ is a countable set, a natural choice for $\K[A]$
is the space $\ell^2(A)$ of square-summable sequences of
complex numbers indexed by the elements of $A$. However,
care is now needed  since the map in \eq{Def:lfva} may
not take a square-summable sequence into one that is also
square summable; for example, this happens if
$f:A\rightarrow B$ is a constant function. In this
Section, we shall restrict our attention to categories of
finite sets, so everything is well defined.

\subsubsection{The Detailed Calculations}
\label{SubSubSec:DetailedCalcs} If $A$ is a finite set,
then any $\vec v\in \mathC^{|A|}$, can be expressed as a
sum
\begin{equation}
       \vec v=\sum_{a\in A}v_a\,\vec a
\end{equation}
where $\vec a$ denotes the element $a\in A$ thought of as
a basis vector for $\mathC^{|A|}$. The expansion
coefficients $v_a\in\mathC$, $a\in A$, are given in terms
of the inner product
$\langle\cdot\;,\cdot\rangle_{\mathC^{|A|}}$ as
\begin{equation}
   v_a=\langle \vec a,\vec v\rangle_{\mathC^{|A|}}.
\end{equation}
In this notation, the linear transformation
$\kappa(f):\mathC^{|B|}\rightarrow \mathC^{|A|}$ in
\eq{Def:lfva}, associated with $f:A\rightarrow B$, is
\begin{equation}
    \kappa(f): \sum_{b\in B} v_b\,\vec b\mapsto
    \sum_{a\in A}v_{f(a)}\,\vec a.   \label{lf}
\end{equation}

For each $b\in\Ran f\subset B$, and for all $a\in
f^{-1}\{b\}$, the value of $v_{f(a)}$ on the right hand
side of \eq{lf} is constant, and equal to $v_b$. Thus, as
$a$ ranges of $f^{-1}\{b\}$ we get a linear sum of the
corresponding basis vectors $\vec a$. This can be turned
into a sum over $b\in B$ by noting that only those $b$
arise that are in the image of $f$. Thus the map in
\eq{lf} can be rewritten as
\begin{equation}
\kappa(f): \sum_{b\in B} v_b\,\vec b\mapsto \sum_{b\in
B}v_b\,\longvec{f^{-1}\{b\}}\label{Def:lf2}
\end{equation}
where
\begin{equation}
    \longvec{f^{-1}\{b\}}:=\sum_{a\in f^{-1}\{b\}}\vec a
\end{equation}
denotes the sum in $\mathC^{|A|}$ of the basis vectors
associated with those elements in $A$ that are mapped by
$f$ to $b\in B$. From this perspective, the fundamental
coherence condition \eq{lambdaConditions} follows from the
basic property of inverse set maps:
\begin{equation}
        f^{-1}(g^{-1}\{c\}) =(g\circ f)^{-1}\{c\}
\end{equation}
for all $f:A\rightarrow B$ and $g:B\rightarrow C$.

Since $\kappa(f): \mathC^{|B|}\rightarrow \mathC^{|A|}$
is linear, it can be written as
\begin{equation}
\kappa(f): \sum_{b\in B} v_b\,\vec b\mapsto
                \sum_{b\in B}v_b\,\kappa(f)\vec b,
\end{equation}
and so another way of specifying $\kappa(f)$ in \eq
{Def:lf2} is in terms of its action on the basis vectors
$\{\vec b\mid b\in B\}$ of $\mathC^{|B|}$:
\begin{equation}
    \kappa(f)\vec b:=\longvec{f^{-1}\{b\}}=
            \sum_{a\in f^{-1}\{b\}}\vec a.
\end{equation}
In particular, the matrix elements of
$\kappa(f):\mathC^{|B|}\rightarrow\mathC^{|A|}$ are
\begin{equation}
     \langle\vec a,\kappa(f)\vec b\,\rangle_{\mathC^{|A|}}
     =\left\{ \begin{array}{ll}
            1 &\mbox{\ if $a\in f^{-1}\{b\}$, {\em i.e.},
             if $b=f(a)$;}         \\
                 0 & \mbox{\ otherwise.}
                 \end{array}
        \right.
\end{equation}
This shows that, as desired, the multiplier $\kappa$
separates arrows that have the same domain and range.

\subsubsection{The Adjoint of $k$}
The linear map $\kappa(f):\K[\Ran f]\rightarrow \K[\Dom
f]$ is used in the definition \eq{Def:a(X)psi} of the
arrow field operator $\hat a(X)$, with
$m(X,A):=\kappa(X(A))$. On the other hand, as shown in
{\bf I.}, the corresponding expression (for finite sets)
for the adjoint $\hat a(X)^\dagger$ is
\begin{equation}
    (\hat a(X)^\dagger\psi)(B)=
\sum_{A\in\ell_X^{-1}\{B\}}\kappa(X(A))^\dagger\psi(A)
\end{equation}
where $\kappa(X(A))^\dagger:\K[A]\rightarrow\K[\ell_XA]$
is the adjoint of the linear map
$\kappa(X(A)):\K[\ell_XA]\rightarrow\K[A]$.

To compute $\kappa(X(A))^\dagger$ explicitly, note that
if $f:A\rightarrow B$, then
$\kappa(f)^\dagger:\mathC^{|A|}\rightarrow \mathC^{|B|}$
is defined by
\begin{equation}
\langle \kappa(f)^\dagger\vec w,\vec
v\,\rangle_{\mathC^{|B|}}:=
    \langle\vec w,\kappa(f)\vec v\,\rangle_{\mathC^{|A|}}
\end{equation}
for all $\vec v\in \mathC^{|B|}$ and $\vec w\in
\mathC^{|A|}$. With respect to the basis $\{\vec b\mid b\in
B\}$ of $\mathC^{|B|}$, we have
\begin{eqnarray}
        (\kappa(f)^\dagger\vec w)_b&=&
            \langle \kappa(f)^\dagger\vec w,\vec
b\,\rangle_{\mathC^{|B|}}=\langle\vec w,\kappa(f)\vec
b\,\rangle_{\mathC^{|A|}}
        =\langle\vec w,\sum_{a\in f^{-1}\{b\}}\vec
        a\,\rangle_{\mathC^{|A|}}\nonumber\\
        &=&\sum_{a\in f^{-1}\{b\}}w_a
\end{eqnarray}
which gives the expression for
$\kappa(f)^\dagger:\mathC^{|A|}\rightarrow \mathC^{|B|}$:
\begin{eqnarray}
\kappa(f)^\dagger\vec w&=&\sum_{b\in B}\sum_{a\in
f^{-1}\{b\}}w_a\vec b=\sum_{b\in B}\sum_{a\in
A}w_a\delta_{b,f(a)}\vec b\nonumber\\
    &=&\sum_{a\in
    A}w_a\longvec{f(a)}.\label{lambdaprime}
\end{eqnarray}
for all $\vec w\in\mathC^{|A|}$.

\subsection{Quantum Measure Theory}
\label{SubSec:QMT}
\subsubsection{The General Idea}
\label{SubSubSec:GenIdeaQMT} A more general way of
associating Hilbert spaces $\K[A]$ with the sets
$A\in\Ob\Q$ is if $\Q$ is a category of {\em measure
spaces\/} in which each set $A\in\Ob\Q$ is equipped with
a measure $\mu_A$ on a $\sigma$-field of subsets of
$A$.\footnote{Note that a scheme of this type allows for
the possibility that the objects in the category may
include more than one copy of the same set equipped with
different measures (and associated fields of measurable
sets) on that set.}  The arrows in this category are
functions $f:A\rightarrow B$ that are measurable with
respect to the measure structures on $A$ and
$B$.\footnote{It may be appropriate to identify functions
$f_1,f_2:A\rightarrow B$ that are equal up to a set of
$\mu_A$-measure zero.} We can then form the bundle of
Hilbert spaces $A\leadsto L^2(A,d\mu_A)$ and try using
this as the quantum bundle $A\leadsto\K[A]$. For example,
if $A$ is a finite set, a natural choice for $\mu_A$ is
$\mu_A(M)=|M|$, for any subset $M$ of $A$. The Hilbert
space $L^2(A,d\mu_A)$ is then isomorphic to
$\mathC^{|A|}$, and we recover the situation discussed in
Section \ref{SubSec:QSTFinite}.

We now proceed as in Section
\ref{SubSubSec:QSTMultipliersFinite} and use
\eq{Def:lfva} to  define the linear map
$\kappa(f):L^2(B,d\mu_B)\rightarrow L^2(A,d\mu_A)$ for
each arrow (measurable map) $f:A\rightarrow B$. However,
care is needed if some of the Hilbert spaces
$L^2(A,d\mu_A)$, $A\in\Ob\Q$, are infinite dimensional
since the map $\kappa(f)$ could then be unbounded. This
is similar to the problem mentioned earlier in the
context of using \eq{Def:lfva} when some of the sets $A$
are countably infinite.

This situation can be tackled in various ways. One is to
allow the maps $\kappa(f)$ to be unbounded, with domains
that are dense subsets of the appropriate Hilbert spaces.
A more attractive scheme is to restrict the arrows in the
category to be measurable maps for which these problems
do not arise. Of course, if such a quantum scheme is to
be applied to a concrete physical problem, there need to
be good physical reasons for making restrictions of this
type.

\subsubsection{The Special Case of a Category of Groups}
\label{SubSubSec:QSTGroups} One situation where measures
arise naturally is if $\Q$ is a small category of locally
compact topological groups, with the arrows being
continuous group homomorphisms. Each such group $G$ has a
unique Haar measure $\mu_G$, and this gives a preferred
choice for the Hilbert spaces $\K[G]$: namely,
$L^2(G,d\mu_G)$.

Then, if $\phi:G\rightarrow H$ is a continuous
homomorphism, we follow the example of \eq{Def:lfva} and
define a linear map
$\kappa(\phi):L^2(H,d\mu_H)\rightarrow L^2(G,d\mu_G)$ by
\begin{equation}
        (\kappa(\phi)v)(g):=v(\phi(g))
\end{equation}
for all $g\in G$ and $v\in L^2(H,d\mu_H)$ (once again,
care is needed if any of the Hilbert spaces have infinite
dimension).

An important property of the Hilbert space $L^2(G,d\mu_G)$ is that
it carries the  regular representation of $G$ defined by
\begin{equation}
            (l_gv)(g_0):=v(g_0g).  \label{Def:RegRepnG}
\end{equation}
Although it is not strictly relevant to our needs, we
note that the maps $\kappa(\phi):L^2(H,d\mu_H)\rightarrow
L^2(G,d\mu_G)$ are actually {\em intertwining\/}
operators for the regular representations on the two
groups. Thus, if $\psi\in L^2(H,d\mu_H)$ and
$\phi:G\rightarrow H$ is any continuous group
homomorphism, we have
\begin{eqnarray}
    (l_g\kappa(\phi)\psi)(g_0)&=&(\kappa(\phi)\psi)(g_0g)
    =\psi[\phi(g_0g)]\nonumber\\
    &=&\psi[\phi(g_0)\phi(g)]=(l_{\phi(g)}\psi)[\phi(g_0)]
    \nonumber\\
    &=&(\kappa(\phi)\ell_{\phi(g)}\psi)(g_0)
\end{eqnarray}
for all $g_0,g\in G$. This gives the intertwining
relations
\begin{equation}
    l_g\kappa(\phi)=\kappa(\phi)l_{\phi(g)}
\end{equation}
corresponding to the commutative diagram
\begin{equation}
        \bundle{L^2(G)}{l_g}{L^2(G)}
        \bundlemapright{\kappa(\phi)}{\kappa(\phi)}
        \bundle{L^2(H)}{l_{\phi(g)}}{L^2(H)}
\end{equation}

\subsection{Categories of Causal Sets or
Topological Spaces}
\subsubsection{Cautionary Remarks}
\label{SubSubSec:CautionaryRemarks} Let us turn now to
the problem of quantising on a category $\Q$ of causal
sets, or topological spaces, or manifolds, etc., in which
the arrows $f:A\rightarrow B$ are structure-preserving
functions. As a first try we could forget about this
extra structure and use the category $\Q_S$ whose objects
are defined to be the same as those of $\Q$ but whose
arrows $f:A\rightarrow B$ are now {\em all\/} functions
from $A$ to $B$, not just those that are structure
preserving.

Suppose that we succeed in constructing a proper
quantisation on $\Q_S$ by finding a representation of the
category quantisation monoid that is both irreducible and
faithful. Then since each arrow in $\Q$ is also a
function between sets, and hence an arrow in $\Q_S$, it
follows that each arrow field in $\AF\Q$ will also belong
to $\AF{\Q_S}$, and will hence be represented in the
quantum theory on $\Q_S$. Furthermore, since the quantum
theory on $\Q_S$ separates arrows, {\em a fortiori\/} the
same will be true for the subset of arrows that belong to
$\Q$. Therefore, the representation of the category
quantisation monoid of $\Q$ obtained in this way will
certainly be faithful.

However, there is no guarantee that this representation
is {\em irreducible\/}. For example, if $A$ is a finite
set, the Hilbert space $\mathC^{|A|}$ carries an
irreducible representation of the monoid $\Map{A,A}$, but
it may be reducible with respect to the submonoid of
order-preserving maps, or continuous maps, etc. If this
happens for at least one object $A\in\Ob\Q$, then the
representation of the category quantisation monoid of
$\Q$ will no longer be irreducible\footnote{It is being
assumed here that $\Q$ has a finite or countable number
of objects with an associated inner product
\eq{ScalProdCount}. The discussion needs to be extended
in an obvious way for the inner product in
\eq{Def:innerproduct} with a general measure $\mu$ on
$\Ob\Q$.}. And even if the representation of $\Hom{A,A}$
{\em is\/} irreducible for all $A\in\Ob\Q$,  the
representation of the category quantisation monoid could
still be reducible if there are not enough arrows between
pairs of {\em different\/} objects.

It might be possible to say something about this issue in
a general way, but probably it will need to be tackled on
a case-by-case basis, with each specific category $\Q$ of
physical interest being examined in detail.

\subsection{A Very Simple Example} \label{SubSec:VSEx}
\subsubsection{Quantisation with State Functions}
We will now illustrate the general scheme with the aid of a very
simple example of a category with just two objects: the causal
sets (posets) $A:=\{a\}$, and $B:=\{b,c\}$ with $b\leq c$. The
arrows are order-preserving maps, and there are two such $f_1,
f_2$ from $A$ to $B$ defined by
\begin{eqnarray}
            f_1(a)&:=&b         \\
            f_2(a)&:=&c
\end{eqnarray}
respectively, and one arrow $g$ from $B$ to $A$ defined by
\begin{equation}
            g(b):=a,\ g(c):=a.
\end{equation}

A non-trivial arrow $r:B\rightarrow B$ is
\begin{equation}
    r(b):=b,\ r(c):=b
\end{equation}
which is not invertible. The same applies to the other
non-trivial arrow $s:B\rightarrow B$ defined by
\begin{equation}
    s(b):=c,\ s(c):=c.
\end{equation}

If we forget the causal/ordering structure on the set $B$, then an
additional arrow $p:B\rightarrow B$ can be defined by
\begin{equation}
p(b):=c,\ p(c):=b           \label{Def:mapp}
\end{equation}
which is the non-trivial element of the permutation group
$\mathZ_2$ of the set $B$. It is not, however, an arrow
in the category $\Q$ since it reverses the ordering of
the elements $b,c\in B$.

In summary, we have the following sets of arrows:
\begin{eqnarray}
        \Hom{A,B}&=&\{f_1,f_2\}     \\
        \Hom{B,A}&=&\{g\}           \\
        \Hom{A,A}&=&\{\id_A\}       \\
        \Hom{B,B}&=&\{\id_B,r,s\}
\end{eqnarray}
to which should be added the map $p:B\rightarrow B$ in
\eq{Def:mapp} if we forget the causal structure on $B$.

According to the simple quantum scheme without multipliers, the
quantum Hilbert space is $\mathC^2$, where $\psi\in\mathC^2$ is
represented by the pair of complex numbers
$(\psi_A,\psi_B):=(\psi(A),\psi(B))$. Then, using the definition
of $\hat a(f):=\hat a(X_f)$ in \eq{aXf}, we find
\begin{eqnarray}
    \hat a(f_1):(\psi_A,\psi_B)&\mapsto&(\psi_B,\psi_B)\\
    \hat a(f_ 2):(\psi_A,\psi_B)&\mapsto&(\psi_B,\psi_B)\\
    \hat a(g):(\psi_A,\psi_B)&\mapsto&(\psi_A,\psi_A)\\
    \hat a(\id_A):(\psi_A,\psi_B)&\mapsto&(\psi_A,\psi_B).
\end{eqnarray}
This shows explicitly the failure to separate the arrows
$f_1:A\rightarrow B$ and $f_2:A\rightarrow B$. Similarly,
the arrows in $\Hom{B,B}=\{\id_B,r,s\}$ are all
represented by the unit operator on $\mathC^2$, and are
hence not separated.

Thus the simple quantum scheme cannot distinguish between
the given category and the one whose objects are two
singleton sets $A:=\{a\}$ and $B:=\{b\}$, with just one
arrow from $A$ to $B$ and one arrow from $B$ to $A$ (plus
$\id_A$ and $\id_B$ as always).

\subsubsection{Quantisation With a Multiplier}

Now consider the quantisation on the same category, but
this time using a multiplier. The discussion in Section
\ref{SubSubSec:DetailedCalcs} of quantising on a category
 of finite sets, suggests that the appropriate choices for the Hilbert space fibres
are $\K[A]=\mathC^1\simeq\mathC$, and $\K[B]=\mathC^2$.
Thus the quantum state space of the system is now
$\mathC\oplus\mathC^2\simeq\mathC^3$; a vector $\psi$ in
this space will be denoted by
$(\psi_A;\psi_{B_1},\psi_{B_2})\in\mathC^3$.

The expression \eq{Def:lf2} for the multiplier, gives the
following representations \eq{Def:a(X)psi} of the
non-trivial arrow fields $X_f$, $f\in\Hom\Q$,
\begin{eqnarray}
\hat a(f_1):(\psi_A;\psi_{B_1},\psi_{B_2})&\mapsto&
            (\psi_{B_1};\psi_{B_1},\psi_{B_2})\label{af1} \\
\hat a(f_2):(\psi_A;\psi_{B_1},\psi_{B_2})&\mapsto&
            (\psi_{B_2};\psi_{B_1},\psi_{B_2})\\
\hat a(g):(\psi_A;\psi_{B_1},\psi_{B_2})&\mapsto&
            (\psi_A;\psi_A,\psi_A)\\
\hat a(r):(\psi_A;\psi_{B_1},\psi_{B_2})&\mapsto&
            (\psi_A;\psi_{B_1},\psi_{B_1})\\
\hat a(s):(\psi_A;\psi_{B_1},\psi_{B_2})&\mapsto&
            (\psi_A;\psi_{B_2},\psi_{B_2}).\label{as}
\end{eqnarray}
Note that arrows with the same domain and range are now
separated, as required.

Of course, it is easy to write these operators as $3\times
3$ matrices. For example,
\begin{equation}
    \hat a(f_1):\pmatrix{a\cr b\cr c}\mapsto
    \pmatrix{b\cr b\cr c}=\pmatrix{0&1&0\cr
    0&1&0\cr 0&0&1}\pmatrix{a\cr b\cr c}
\end{equation}
so that
\begin{equation}
    \hat a(f_1)=
    \pmatrix{0&1&0\cr 0&1&0\cr 0&0&1}
\end{equation}
with the adjoint
\begin{equation}
    \hat a(f_1)^\dagger=
    \pmatrix{0&0&0\cr 1&1&0\cr 0&0&1}
\end{equation}

If we forget the causal structure on the set $B$, then
there is the additional arrow $p:B\rightarrow B$ defined in
\eq{Def:mapp}. The corresponding arrow field $X_p$ is
represented in the quantum theory by
\begin{equation}
\hat a(p):(\psi_A;\psi_{B_1},\psi_{B_2})\mapsto
            (\psi_A;\psi_{B_2},\psi_{B_1})\label{ap}
\end{equation}

\section{Further Quantum Operations}
\label{Sec:FurtherQuOps} \subsection{Combining Causal Sets}
\label{SubSec:CombiningCausalSets}
 There
are some special operations on causal sets (posets) that one might
wish to incorporate in the present scheme. For example
\cite{DP90}:
\begin{enumerate}

\item If $A$ and $B$ are disjoint posets, the {\em
disjoint union\/} $A\cup B$ is defined on the set-theoretic
disjoint union by the ordering $x\leq y$ if
\begin{enumerate}
       \item $x,y\in A$ and $x\leq_A y$; {\em or}
       \item $x,y\in B$ and $x\leq_B y$.
\end{enumerate}

\item If $A$ and $B$ are disjoint posets,  the
{\em linear sum\/} $A\oplus B$ is defined on the
set-theoretic disjoint union with the ordering $x\leq y$ if
\begin{enumerate}
       \item $x,y\in A$ and $x\leq_A y$; {\em or}
       \item $x,y\in B$ and $x\leq_B y$; {\em or}
       \item $x\in A$ and $y\in B$.
\end{enumerate}

\item If $A$ and $B$ are posets, the set-theoretic
Cartesian product $A\times B$ can be
given a poset structure by defining $(x,y)\leq (p,q)$ if
$x\leq_A p$ and $y\leq_B q$.
\end{enumerate}
Of course, these constructions are only of use if the
category concerned includes them. For example, the
disjoint union of two connected posets is disconnected,
and therefore would not be included if $\Q$ is a category
of connected causal sets. On the other hand, the linear
sum of connected posets {\em is\/} connected, and
therefore could be used.

Let us start with the linear sum. In a simple
quantisation scheme with complex-valued state functions
on $\Ob\Q$, it is natural to define operators $\hat D_B$,
$B\in\Ob\Q$, by
\begin{equation}
    (\hat D_B\psi)(A):=\psi(A\oplus B)\label{Def:DBSimple}
\end{equation}
for all $A\in\Ob\Q$. These represent the $\oplus$ operation
in the sense that $\hat D_C\hat D_B=\hat D_{C\oplus B}$.

Things are more complicated when multipliers are used
since it is necessary to construct an appropriate linear
map from $\K[A\oplus B]$ to $\K[A]$. To do this within
the category quantisation scheme we need to pick some
preferred arrow from $A$ to $A\oplus B$. The obvious
choice is the injection $i_{A,A\oplus B}:A\rightarrow
A\oplus B$, which leads to the definition
\begin{equation}
    (\hat D_B\psi)(A):=\kappa(i_{A,A\oplus B})
    \psi(A\oplus B). \label{Def:DB}
\end{equation}
Of course, the same thing can be done for the disjoint
union: {\em i.e.}, use the natural embedding of $A$ in
$A\cup B$.

For the product operation, the analogy of \eq{Def:DBSimple}
is
\begin{equation}
    (\hat \Pi_B\psi)(A):=\psi(A\times B).
\end{equation}
However, the analogue of \eq{Def:DB} needs more thought.
This is because the natural map associated with a product
$A\times B$ is the projection $\pi_{A\times B, A}:A\times
B\rightarrow A$, $(a,b)\mapsto a$. But the linear map
$\kappa(\pi_{A\times B, A})$ is then from $\K[A]$ to
$\K[A\times B]$, which is in the wrong direction. The
obvious resolution is to use the adjoint
$\kappa(\pi_{A\times B, A})^\dagger$, which maps
$\K[A\times B]$ to $\K[A]$. Thus we define
\begin{equation}
(\hat \Pi_B\psi)(A):= \kappa(\pi_{A\times B, A})^\dagger
\psi(A\times B).       \label{Def:PiB}
\end{equation}
By these means, the causal-set operations of direct sum,
disjoint union, and product can be incorporated into the
category quantisation scheme.

Clearly, in a general category of small sets $\Q$,
analogues of \eq{Def:DB} and \eq{Def:PiB} can be used to
represent the operations of taking, respectively, the
disjoint union, and product, of two sets providing these
exist as objects in $\Q$.

\subsection{When $\Ob\Q$ has the Structure of a Lattice} A
mathematically similar situation to the poset operations
in Section \ref{SubSec:CombiningCausalSets} arises in the
context of an older approach to quantum topology
discussed in \cite{Ish89}. This involved quantising the
set of topologies $\tau(X)$ on a fixed set $X$ by
exploiting the fact that $\tau(X)$ is an algebraic {\em
lattice\/}.\footnote{Within the context of our present
discussion, this earlier scheme is deficient in the sense
that multipliers were not used, and hence the internal
structure of each topology $\tau\in\tau(X)$ is not
represented. It would be interesting to revisit that
earlier work in the light of the more sophisticated
category quantisation techniques.} In general, if the set
of objects $\Ob\Q$ of a small category $\Q$ is a lattice,
then the `$\lor$' and `$\land$' lattice operations can be
incorporated into the category quantisation scheme in
direct analogy with equations \eq{Def:DB} and
\eq{Def:PiB}. Thus, for the $\lor$ operation we exploit
the canonical arrow $o_{A,A\lor B}:A\rightarrow A\lor B$
(where $o_{x,y}:x\rightarrow y$ is the arrow associated
with $x\leq y$) and define
\begin{equation}
    (\hat O_B\psi)(A):=\kappa(o_{A,A\lor B})
    \psi(A\lor B),\label{Def:OBpsiA}
\end{equation}
while, for the $\land$ operation, we use the arrow
$o_{A\land B,A}:A\land B\rightarrow A$ and define
\begin{equation}
(\hat \Pi_B\psi)(A):= \kappa(o_{A\land B, A})^\dagger
\psi(A\land B).\label{Def:PiBpsiA}
\end{equation}

It is easy to check that, by these means, the lattice
operations are represented in the quantum theory in the
sense that
\begin{equation}
    \hat O_B\hat O_C=\hat O_{B\lor C}
\end{equation}
and
\begin{equation}
    \hat \Pi_B\hat \Pi_C=\hat \Pi_{B\land C}.
\end{equation}

A special case is the lattice of subsets of some master
set $U$. The quantisation scheme discussed in Section
\ref{Sec:QST} can then be augmented with operations
\begin{eqnarray}
    (\hat O_B\psi)(A)&:=&\kappa(o_{A,A\cup B})
    \psi(A\cup B)\\
    (\hat \Pi_B\psi)(A)&:= &\kappa(o_{A\cap B, A})^\dagger
\psi(A\cap B)
\end{eqnarray}
for all $A,B\subset U$, which are the analogues of
\eq{Def:OBpsiA} and \eq{Def:PiBpsiA} respectively. In a
similar way, one can implement the operation of taking
the complement of a set in which $A\mapsto U-A$.

\section{Quantum Field Theory in this Language}
\label{Sec:QFT} Standard quantum field theory can also be
discussed in the language of quantising on a category, and
provides another example (albeit, somewhat contrived) of
how the formalism works.

We start by considering what is superficially the same
category as in \eq{DiagramN}, but with the object $n$ now
being identified with the vector space $\mathR^{3n}$.
This gives the chain
\begin{equation}
    \{0\}\shortmapright{i_0}\mathR^3\shortmapright{i_1}
    \mathR^6\rightarrow
    \cdots\rightarrow\mathR^{3(n-1)}\rightarrow\cdots
\end{equation}
where $i_k:\mathR^{3k}\rightarrow\mathR^{3(k+1)}$ is the
natural embedding of vector spaces
\begin{eqnarray}
            \mathR^{3k}&\shortmapright{i_k}&\mathR^{3(k+1)}
            \label{Def:iotak}\\
            (x_1,x_2,\ldots,x_{3k})&\mapsto&
            (x_1,x_2,\ldots,x_{3k},0,0,0)\nonumber
\end{eqnarray}

If the corresponding quantum theory is viewed as being of a
system whose configuration space can be any these vector
spaces, then each $\mathR^{3k}$ has the Euclidean group
$E(3k)$ as its group of isometries. As such, it is a
natural choice for the monoid of arrows from the object
$\mathR^{3k}$ to itself. Thus once again we have a
situation in which the arrows in the original category can
be extended by including appropriate transformations of the
internal structures of the objects. Thus, in this example
it is reasonable to define the most general arrow from
$\mathR^{3k}$ to $\mathR^{3(k+1)}$ to be of the form
$\beta\circ i_k\circ\alpha$ where $\alpha\in E(3k)$ and
$\beta\in E(3(k+1))$.

The monoids (actually, groups)
$\Hom{\mathR^{3k},\mathR^{3k}}\simeq E(3k)$ can
represented in the quantum theory by associating to each
vector space $\mathR^{3k}$, $k\geq 1$, the Hilbert space
$L^2(\mathR^{3k})$, and associating to the null vector
space $\{0\}$ the space $\mathC$. This Hilbert bundle can
be turned into a contravariant presheaf by defining the
linear operators
$\kappa(i_k):L^2(\mathR^{3(k+1)})\rightarrow
L^2(\mathR^{3k})$ as
\begin{equation}
   (\kappa(i_k)\phi)(x_1,x_2,\ldots,x_k):=
   \phi(x_1,x_2,\ldots,x_k,0,0,0)
\end{equation}
for all $\phi\in L^2(\mathR^{3(k+1)})$. More generally,
the arrow $\beta\circ
i_k\circ\alpha:\mathR^{3k}\rightarrow\mathR^{3(k+1)}$ is
associated with the multiplier
$R^{3k}(\alpha)\kappa(i_k)R^{3(k+1)}(\beta):L^2(\mathR^{3(k+1)})\rightarrow
L^2(\mathR^{3k})$ where $R^{3(k+1)}(\beta)$ denotes the
representation of $\beta\in E(3(k+1))$ on
$L^2(\mathR^{3(k+1)})$, and $R^{3k}(\alpha)$ denotes the
representation of $\alpha\in E(3k)$ on
$L^2(\mathR^{3k})$.

A cross-section of this Hilbert bundle corresponds to an
element of bosonic Fock space, defined as $\mathC\oplus
L^2(\mathR^3)\oplus L^2(\mathR^6)\oplus\cdots$, and so we
are close to recovering non-relativistic, bosonic quantum
field theory. However, in this category approach there is
no obvious way of deriving the usual annihilation and
creation operators of quantum field theory. The problem
is that this particular category does not have enough
arrows between different objects.

This suggests to try again using the chain \eq{DiagramN}
$$
    0\rightarrow 1 \rightarrow 2 \rightarrow\cdots
    \rightarrow n-1 \rightarrow\cdots
$$
but now define the arrows from an object $k$ to $l$
($l>k$) to be functions in $L^2(\mathR^{3(l-k)})$, {\em
i.e.}, $\Hom{k,l}:=L^2(\mathR^{3(l-k)})$. Composition of
arrows/functions is then defined as the tensor product
operation. More precisely, if $f:j\rightarrow k$ and
$g:k\rightarrow l$, so that $f\in L^2(\mathR^{3(k-j)})$
and $g\in L^2(\mathR^{3(l-k)})$, then $g\circ
f:j\rightarrow l$ is defined as $g\circ f:=g\otimes f\in
L^2(\mathR^{3(l-k)})\otimes L^2(\mathR^{3(k-j)})\simeq
L^2(\mathR^{3(l-j)})$. This arrow composition law is
associative since $f\otimes(g\otimes h)=(f\otimes
g)\otimes h$ for functions $f,g,h$. When $k=l$, it is
consistent to identify $L^2(\mathR^{3(l-k)})$ with the
vector space $\mathC$; {\em i.e.}, for all $n\geq 0$, we
set $\Hom{n,n}:=\mathC$, with composition of complex
numbers being defined as multiplication.

Multipliers are needed to separate the arrows between
different objects, and the first step is to construct the
bundle of Hilbert spaces over $\Ob\Q=\mathN$ in which the
fibre $\K[n]$ over $n\in\Ob\Q$ is $L^2(\mathR^{3n})$ for
$n>0$, and with $\K[0]\simeq \mathC$. Then, if
$f:k\rightarrow l$, so that $f\in L^2(\mathR^{3(l-k)})$,
we define $\delta(f):\K[k]\rightarrow \K[l]$ by
$\delta(f)(h):=f\otimes h$, for all $h\in \K[k]\simeq
L^2(\mathR^{3k})$. Note that $f\in L^2(\mathR^3)$ defines
an arrow $f:k\rightarrow k+1$ for all $k>0$, and the
associated (covariant) multiplier
$\delta(f):L^2(\mathR^{3k})\rightarrow
L^2(\mathR^{3(k+1)})$ is the usual Fock space creation
operator $\hat b(f)^\dagger$.

By these means we have constructed a covariant presheaf,
rather than the contravariant one used earlier. A
contravariant presheaf can be obtained by defining the
multiplier $\kappa(f): \K[l]\rightarrow \K[k]$,
$f:k\rightarrow l$, as the adjoint of $\delta(f)$. This
can then be used to define the operators $\hat a(X)$ as
in \eq{Def:a(X)psi}.

Alternatively, the adjoint $\hat a(X)^\dagger$ can be
defined directly using $\delta$. Specifically, as shown
in {\bf I.}, the general expression for the adjoint
operator $\hat a(X)^\dagger$ is
\begin{equation}
    (\hat
    a(X)^\dagger\psi)(B)=\sum_{A\in\ell_X^{-1}\{B\}}
        m(X,A)^\dagger\psi(A)
\end{equation}
for all $A\in\Ob\Q$. In the present case, this becomes
\begin{eqnarray}
    (\hat
    a(X)^\dagger\psi)(n)&=&\sum_{k\in\ell_X^{-1}\{n\}}
        \delta(X(k))\psi(k)\\
        &=&\sum_{\{k\mid n=\ell_Xk\}}X(k)\otimes\psi(k)
\end{eqnarray}

In particular, if $f\in L^2(\mathR^3)$, and if we define
the arrow field $X^f(n):=f:n\rightarrow n+1$ for all
$n\geq 0$, then
\begin{eqnarray}
(\hat a(X^f)^\dagger\psi)(n)&=&f\otimes\psi(n-1)\\
        &=&\hat b(f)^\dagger\psi(n-1)
\end{eqnarray}
for all $n\geq 1$.

 Note that although the objects have no internal
structure, it is still of course true that the Hilbert
space $L^2(\mathR^{3n})$ carries a representation of the
Euclidean group $E(3n$) (and, indeed, of the canonical
commutation relations) in $3n$ dimensions.

\section{Conclusions}
We have seen how the general theory of quantising on a
category $\Q$ can be applied to two different situations.
The first is when the category is a poset, in which case
the objects have no internal structure, and a quantisation
with complex-valued state functions is sufficient.

The second, and more important, situation was one of the original
motivations for studying the category quantisation scheme: namely,
quantisation on a category whose objects are sets that could be
potential models for space-time (or space). In particular, we have
shown how to construct suitable multipliers for a category of
finite sets. We have also indicated how this scheme might be
extended to arbitrary sets equipped with measures.

A major task for future research is to return to the examples of
sets with structure---such as posets, or topological spaces---to
get a better idea of when the irreducible representations of the
category quantisation monoid of the underlying category of sets
remain irreducible when the arrows are restricted to be
structure-preserving maps. It is not clear how much can be said
about this issue in general, and how much will depend on  the
details of the specific category. A useful step in this direction
would be to study a variety of simple models of quantum casual
sets and quantum topologies.

Another important topic for the future is the addition of matter
degrees of freedom to a quantum causal set theory. In doing so, it
will be necessary to amalgamate the insights into the local
Hilbert spaces $\K[A]$, $A\in\Ob\Q$, obtained from the present
paper and {\bf I.}, with the understanding of the nature of
quantum history theory as given by the `history projection
operator' (HPO) approach to consistent history theory. In the HPO
approach, to any causal set $A$ there would be associated a
certain Hilbert space ${\cal J}(A)$ whose projection operators
represent propositions about the matter variables in the
background space-time $A$. The task is to combine these Hilbert
spaces ${\cal J}(A)$ with the Hilbert spaces $\K[A]$ of the
quantum causal set theory in an appropriate way. The resulting
quantum history theory will then involve propositions about the
matter fields {\em and\/} the space-times in which they propagate.

Alternatively, one could start with a canonical approach
to quantising on a causal background---for example, as
discussed in \cite{HMS03}---and then use the techniques
in the present paper to allow the background itself to be
`quantised'.

In either case, it must be emphasised that the techniques
developed in the present paper constitute only a `toolkit' for
constructing such theories. The hard task is to use these tools to
build physically realistic models of `quantum space' or `quantum
space-time'.

\section*{Acknowledgements}

\noindent Support by the EPSRC in form of grant GR/R36572
is gratefully acknowledged.


\begin{thebibliography}{1}

\bibitem{IshQCT1_03} C.~J.~Isham,
\newblock A new approach to quantising space-time:
I.\ Quantising on a general category.
\newblock {\em Adv.\ Theor.\ Phys.} {\bf 7} 331--367 (2003).
\newblock gr-qc/0303060.

\bibitem{IshQCT3_03} C.~J.~Isham,
\newblock A new approach to quantising systems:
III.~State vectors as functions on arrows.
\newblock (2003).

\bibitem{DP90} B.~A.~Davey and H.~A.~Priestley,
\newblock {\em Introduction to Lattices and Order}.
\newblock Cambridge University Press, Cambridge, (1990).

\bibitem{Ish89} C.~J.~Isham,
\newblock Quantum topology and quantization on the lattice of
topologies.
\newblock {\em Class.\ Qu.\ Grav.} {\bf 6},
1509--1534, (1989).

\bibitem{HMS03} L.~Hawkins., F.~Markopoulou and
H.~Sahlmann
\newblock Evolution in quantum causal histories.
\newblock hep-th/0302111, (2003).

\end{thebibliography}
\end{document}